\newif\ifTwoColumn
\newif\ifTechReport

\TechReporttrue    

\ifTwoColumn
	\documentclass[twocolumn]{autart}
	\usepackage{cite}
	
\else
	\ifTechReport
		\documentclass[eview,3p,number,11pt]{elsarticle_arXiv}
		\usepackage{amsthm}
		\usepackage{natbib}
		\newenvironment{pf}{\begin{proof}}{\end{proof}}
		\usepackage{parskip}
		\usepackage{geometry}
		
	\else
		\documentclass[onecolumn]{autart}
		\usepackage{cite}
	\fi
\fi

\usepackage[dvipsnames]{xcolor}

\usepackage{amsmath,amssymb,amsfonts,mathtools,bbm,mathrsfs}
\usepackage{dsfont}
\usepackage[acronym,shortcuts]{glossaries}
\usepackage{glossaries-prefix}
\usepackage{balance}
\usepackage{url}
\usepackage[ruled]{algorithm}
\usepackage{nicefrac}
\usepackage{multirow}
\usepackage{graphicx,subcaption}
\usepackage{booktabs}
\usepackage{tikz}
\usepackage{ulem}
\usepackage{accents}
\usepackage{graphicx}
\usepackage{epstopdf}
\graphicspath{{./figures/}}
\usepackage{marginnote}
\usepackage[shortlabels]{enumitem}
\usepackage{makecell}

\newcommand{\continuanceref}{}
\newtheorem{theorem}{Theorem}

\newtheorem{proposition}{Proposition}
\newtheorem{lemma}{Lemma}
\newtheorem{corollary}{Corollary}

\newtheorem{remark}{Remark}
\newtheorem{condition}{Condition}
\newtheorem{standing}{Standing Assumption}

\usepackage[colorinlistoftodos]{todonotes}

\newcommand{\R}{\mathbb{R}}

\newcommand{\B}{\mathbb{B}}

\newcommand{\N}{\mathbb{N}}

\newcommand{\mc}{\mathcal}

\newcommand{\A}{\mc{A}}

\newcommand{\bbS}{\mathbb{S}}

\newcommand{\diag}{\mathrm{diag}}

\newcommand{\col}{\textrm{col}}

\newcommand{\bs}{\boldsymbol}
\newcommand{\bsone}{\boldsymbol{1}}

\definecolor{olivegreen}{RGB}{39,128,0}

\newcommand\oprocendsymbol{\hbox{$\square$}}
\newcommand\oprocend{\relax\ifmmode\else\unskip\hfill\fi\oprocendsymbol}

\newcommand{\eqdef}{\coloneqq}
\newcommand{\reqdef}{\eqqcolon}

\DeclareMathSizes{10}{10}{6}{4}

\newcommand{\normaltext}[1]{\textnormal{#1}}

\makeglossaries
\newacronym[
prefixfirst={a\ },
prefix={an\ }
]{LP}{LP}{linear program}
\newacronym[
prefixfirst={a\ },
prefix={an\ }
]{LTI}{LTI}{linear time-invariant}
\newacronym{CLF}{CLF}{control Lyapunov function}
\newacronym{MPC}{MPC}{model predictive control}
\newacronym{eMPC}{eMPC}{explicit model predictive control}
\newacronym[
prefixfirst={a\ },
prefix={an\ }
]{LQR}{LQR}{linear quadratic regulator}
\newacronym{mp-QP}{mp-QP}{multi-parametric quadratic program}
\newacronym{mp-LP}{mp-LP}{multi-parametric linear program}
\newacronym{QP}{QP}{quadratic program}
\newacronym[
prefixfirst={a\ },
prefix={an\ }
]{LICQ}{LICQ}{linear independence constraint qualification}
\newacronym[
prefixfirst={a\ },
prefix={an\ }
]{LMI}{LMI}{linear matrix inequality}
\newacronym{PWA}{PWA}{piecewise-affine}
\newacronym{PWA-NN}{PWA-NN}{piecewise-affine neural network}
\newacronym{SDP}{SDP}{semidefinite program}
\newacronym{MIP}{MIP}{mixed-integer program}
\newacronym{ReLU}{ReLU}{rectified linear unit}
\newacronym{PL-NN}{PL-NN}{piecewise linear neural network}
\newacronym{PARC}{PARC}{piecewise-affine regression and classification}
\newacronym{NN}{NN}{neural network}
\newacronym{iid}{i.i.d.}{independent and identically distributed}
\newacronym{wrt}{w.r.t.}{with respect to}
\newacronym{RMS}{RMS}{root mean square}
\newacronym{BFR}{BFR}{best fit ratio}
\newacronym[
prefixfirst={a\ },
prefix={an\ }
]{MI}{MI}{mixed-integer}
\newacronym[
prefixfirst={a\ },
prefix={an\ }
]{MILP}{MILP}{mixed-integer linear program}
\newacronym{LIQC}{LIQC}{linear independence constraint qualification}
\newacronym{KKT}{KKT}{Karush-Kuhn-Tucker}
\newacronym{ISS}{ISS}{input-to-state stable}
\newacronym{LCP}{LCP}{linear complementarity problem}
\newacronym{LC}{LC}{linear complementarity}
\newacronym{MLD}{MLD}{mixed-logical dynamical}
\newacronym{MIQP}{MIQP}{mixed-integer quadratic program}
\newacronym{OCP}{OCP}{optimal control problem}
\newacronym{MPCC}{MPCC}{mathematical program with complementarity constraints}
\newacronym{NLP}{NLP}{nonlinear programming}

\begin{document}
\normalem

\begin{frontmatter}

\title{A neural network-based approach to hybrid systems\\ identification for control}

\author{Filippo Fabiani$^\dagger$, Bartolomeo Stellato$^\ddagger$, Daniele Masti$^\star$ and Paul J. Goulart$^\clubsuit$}

\address{$^\dagger$ IMT School for Advanced Studies Lucca, Piazza San Francesco 19, 55100, Lucca, Italy\\$^\ddagger$ Operations Research and Financial Engineering, Princeton University, Princeton, NJ, USA\\
$^\star$ Gran Sasso Science Institute, Viale F. Crispi 7, 67100, L'Aquila, Italy\\$^\clubsuit$ Department of Engineering Science, University of Oxford, OX1 3PJ, Oxford, United Kingdom}
\thanks{\emph{Email addresses:} \texttt{filippo.fabiani@imtlucca.it}, \texttt{bstellato@princeton.edu}, \texttt{daniele.masti@gssi.it}, \texttt{paul.goulart@eng.ox.ac.uk}.}

\begin{abstract}
	We consider the problem of designing a machine learning-based model of an unknown dynamical system from a finite number of (state-input)-successor state data points, such that the model obtained is also suitable for optimal control design.  We adopt a \gls{NN} architecture that, once suitably trained, yields a hybrid system with continuous \gls{PWA} dynamics that is differentiable with respect to the network's parameters, thereby enabling the use of derivative-based training procedures. We show that a careful choice of our \gls{NN}'s weights produces a hybrid system model with structural properties that are highly favorable when used as part of a finite horizon \gls{OCP}.   Specifically, we rely on available results to establish that optimal solutions with strong local optimality guarantees can be computed via \gls{NLP}, in contrast to classical \glspl{OCP} for general hybrid systems which typically require mixed-integer optimization. Besides being well-suited for optimal control design, numerical simulations illustrate that our \gls{NN}-based technique enjoys very similar performance to state-of-the-art system identification methods for hybrid systems and it is competitive on nonlinear benchmarks.
\end{abstract}

\end{frontmatter}

\section{Introduction}\label{sec:intro}

Within the systems-and-control community, \emph{system identification} methods have greatly benefited from powerful tools originating in the machine learning literature \cite{pillonetto2014kernel,breschi2016piecewise,bemporad2023piecewise}. In particular, \glspl{NN} \cite{Goodfellow-et-al-2016} have been used since the 1980s \cite{werbos1989neural} to produce data-driven system model surrogates.
Their flexibility—stemming from their universal approximation capability \cite{hornik1989multilayer} and the wide range of available architectures—has motivated research into {NN}-based techniques for (non)linear system identification adopting various structures \cite{forgione2021continuous,forgione2023adaptation,andersson2019deep,mavkov2020integrated,masti2021learning}.

Despite a few exceptions \cite{hoekstra2023computationally,liu1998predictive}, available learning-based system identification techniques do not consider whether the resulting models possess an internal structure favourable for use within an \gls{OCP}.  In fact, unless one assumes a certain simplified structure, deep \glspl{NN} are generally hard to analyze due to their nonlinear and large-scale structure \cite{Goodfellow-et-al-2016}. As an undesirable consequence, adopting the resulting system model surrogates into \glspl{OCP} may make the latter extremely hard to solve efficiently (wherever a solution is guaranteed to exist).   In this paper we propose a \gls{NN}-based technique to identify a hybrid system representation, in the form of continuous \gls{PWA} dynamics, with specific structure suitable for optimal control design.

\subsection{Related work}
Although hybrid models represent a broad class of systems \cite{garulli2012survey} whose identification is known to be NP-hard in general \cite{roll2004identification,lauer2015complexity}, to the best of our knowledge little attention has  focused on the derivation of such models through (deep) \gls{NN}-based surrogates. 

Starting from the pioneering work in \cite{batruni1991multilayer}, where a continuous \gls{PWA} representation was obtained through a \gls{NN} with a specific two-layer structure by relying on \cite{chua1988canonical}, \cite{choi1994constructive} constructed a \gls{PWA} local model by introducing an input space tessellation via hyperpolyhedral convex cells, associating to them \gls{NN} granules with a local interpolation capability. In \cite{hush1998efficient}, instead, the authors introduced a constructive algorithm that builds a \gls{NN} with \gls{PWA} sigmoidal nodes and one hidden layer. In particular, one node at a time was added by fitting the residual, a task that was accomplished by searching for the best fit through a sequence of \glspl{QP}. A single hidden layer network was also proposed in \cite{gad2000algorithm}, which constructed a continuous \gls{PWA} error function and developed an efficient, two-step algorithm working in the weight space to minimize it. Recently, \cite{falt2019identification} proposed a series of experiments in which novel libraries are employed to identify dynamical models with \glspl{NN} for complicated hybrid systems, while \cite{de2018end,li2018learning} have analyzed the effects of exploiting structured knowledge to \gls{NN} surrogates in describing the system multi-modal behaviour.

Similar to our approach, \cite{jin2022learning} proposed to infer a control-oriented \gls{LC} dynamics (equivalent to a \gls{PWA} one \cite{heemels2001equivalence}) directly from data. In particular, the approach in \cite{jin2022learning} built upon the minimization of a tailored violation-based loss function, which allowed one to learn \gls{LC} dynamics via standard gradient-based methods. In view of our results, we also mention the approach in \cite{yang2022modeling}. There, a stationarity condition was identified as necessary and sufficient to characterize local optima of a program with a \gls{ReLU} \gls{NN} entering both in the cost and constraints, and the corresponding \gls{LC}-based reformulation yielding an \gls{MPCC}.

\subsection{Summary of contribution}
	Following our recent contributions \cite{fabiani2023reliably,fabiani2022neural}, in this paper we take a \gls{NN}-based approach to the problem of system identification for control. In particular, we wish to answer the following question: given $N$ (state-input)-successor state samples $\{(x^{(i)},u^{(i)}, x^{+,{(i)}})\}_{i=1}^N$, how can we obtain a descriptive system model that is also suitable for optimal control design? While the meaning of ``descriptive'' and ``suitable'' will be clarified later in the paper, we summarize the contribution as follows:
	\begin{enumerate}
		\item We employ a \gls{NN}-based method to obtain a hybrid system with continuous \gls{PWA} dynamics from available data. The simple \gls{NN} architecture we adopt combines an OptNet layer \cite{amos2017optnet,amos2018differentiable} and an affine one, and results in a differentiable output \gls{wrt} the \gls{NN}'s parameters;
		\item Given its end-to-end trainability, we show that a careful choice of some weights of the \gls{NN} allows us to produce a hybrid dynamics model with a specific structure. By relying on the results in \cite{hempel2017strong}, we then establish that the latter can be controlled (locally) optimally by solving the \gls{KKT} conditions of the underlying \gls{OCP};
		\item Extensive numerical simulations show that, as a \gls{NN}-based identification procedure, our technique has very similar performance compared to the state-of-the-art of hybrid system identification methods.
	\end{enumerate}

	Our method thus requires a simple identification step, represented by a careful training of a \gls{NN} with specific structure via standard tools, which yields a hybrid system model that is well-suited to optimal control design. In particular, solving an \gls{OCP} involving a hybrid system with \gls{PWA} dynamics as an \gls{NLP} has been proven to require shorter computation times and feature better scaling in the problem dimensions than standard approaches based on mixed-integer optimization \cite{hempel2017strong,hall2021sequential}. In addition, we recall that \gls{PWA} regression is NP-hard in general \cite{lauer2015complexity}, since it requires simultaneous classification of the $N$ samples $\{(x^{(i)},u^{(i)}, x^{+,{(i)}})\}_{i=1}^N$ into modes, thereby calling for a regression of a submodel for each mode. 
	By taking a \gls{NN}-based perspective, we circumvent such a potentially challenging classification issue, thereby reducing the identification step to the training of the adopted \gls{NN} architecture, a task that can be accomplished through standard gradient-based methods in view of the network's output differentiability.
	
	Remarkably, we obtain an \gls{LC} model suitable for control as a direct consequence of a careful choice of a \gls{NN} with specific architecture, thereby circumventing the requirement of strict complementarity to recover differentiability \gls{wrt} the main parameters as in \cite{jin2022learning,de2018end}. In contrast to \cite{yang2022modeling}, we give an explicit structure of a \gls{NN} for which local stationarity conditions, coinciding with the standard \gls{KKT} system, are known to hold for the \gls{MPCC} obtained by  embedding the \gls{NN} as a hybrid model in an \gls{OCP}.

	The rest of the paper is organized as follows: we formalize the problem addressed in \S \ref{sec:problem_description}, and in \S \ref{sec:neural_network_representation} introduce our \gls{NN}-based approach to system identification for control. In \S \ref{sec:OCP} we describe the \gls{OCP} obtained when using our identified model structure, and develop our main theoretical results relating to that problem in \S \ref{sec:main_result}. Finally, \S \ref{sec:simulations} reports a number of extensive numerical experiments.

\subsection*{Notation}
$\N$, $\R$ and $\R_{\geq 0}$ denote the set of natural, real and nonnegative real numbers, respectively.
$\bbS^{n}$ is the space of $n \times n$ symmetric matrices and $\bbS_{\succ 0}^{n}$ ($\bbS_{\succcurlyeq 0}^{n}$) is the cone of positive (semi-)definite matrices. Bold $\bsone$ ($\bs{0}$) is a vector of ones (zeros). Given a matrix $A \in \R^{m \times n}$,
$A^\top$  denotes its transpose. 
For $A \in \bbS_{\succ 0}^{n}$, $\| v \|_A \eqdef \sqrt{ v^\top A v }$. The operator $\col(\cdot)$ stacks its arguments in column vectors or matrices of compatible dimensions. $A \otimes B$ is the Kronecker product of $A$ and $B$. We sometimes use $x_{k+1}$, $k \in \N_0$, as opposed to $x^+$, to make time dependence explicit when describing the state evolution of discrete-time systems.
The uniform distribution on the interval $[a,b]$ is denoted by $U(a,b)$, whereas $N(\rho, \sigma^2)$ stands for the normal distribution with mean $\rho$ and standard deviation $\sigma$.

%

\glsresetall

\section{Problem formulation}
\label{sec:problem_description}
We will assume that we have available a finite collection of (state-input)-successor state measured triplets, $(x,u, x^{+})$, $x$, $x^+\in\R^n$, $u\in\R^m$ for an unknown but deterministic dynamic system.  Our aim is to produce a data-driven model of this unknown system without running further experiments.
Such a model shall be ``descriptive enough'', i.e., it shall belong to a model class capable of capturing a wide variety of system behaviors, while at the same time being suited for optimal control. 

To this end, we will consider throughout the paper the following finite horizon \gls{OCP}:
\begin{equation}\label{eq:OCP}
	\left.
	\begin{aligned}
		&\underset{\bs{u}, \bs{x}}{\textrm{min}} && \ell_T\left(x_T\right)+\sum_{t\in\mc T} \ell_t\left(x_t, u_t\right)\\
		& \textrm{ s.t. } && x_{t+1} = \mc N_\theta(x_t, u_t), \text{ for all } t \in \mc T,\\
		&&& x_0 = x(0),
	\end{aligned}
	\right.
\end{equation}
where $x_t$, $t \in \mc T \eqdef \{0, \ldots, T-1\}$, denotes the predicted system state $t$ timesteps into the future through the (possibly nonlinear) dynamical model $\mc N_\theta : \R^p \times \R^n \times \R^m \to \R^n$, which is to be inferred from data and is characterized by parameters $\theta \in \R^p$. We denote by $\bs u\eqdef\col((u_t)_{t \in \mc T}) \in \R^{mT}$ the collection of control inputs to be chosen and by $\bs x\eqdef\col((x_t)_{t \in \mc T \cup \{T\}}) \in \R^{n(T+1)}$ the resulting state trajectory starting from the measured initial state $x_0=x(0)$. In addition, let $\ell_t$ and $\ell_T$ denote the stage and terminal costs respectively, which are assumed to meet the following standard condition:
\begin{standing}\label{standing:cost}
	The stage cost $\ell_t : \R^n \times \R^m \to \R$ and the terminal cost $\ell_T: \R^n \to \R$ are convex functions.
\end{standing}
%

Thus, given $N$ samples $\{(x^{(i)},u^{(i)}, x^{+,{(i)}})\}_{i=1}^N$, we aim at developing a system model $\mc N_\theta : \R^p \times  \R^n \times \R^m \to \R^n$ such that, once  $\mc N_\theta$ is used within the \gls{OCP} in \eqref{eq:OCP}, an optimal solution to the latter can be characterized via standard optimality conditions. To this end, we design $\mc N_\theta$ as a \gls{NN}-based representation of a hybrid system, ultimately leading to an \gls{OCP} equivalent to \eqref{eq:OCP} in the form of a \gls{MPCC} \cite{luo1996mathematical}, whose \gls{KKT} conditions \cite[\S 5.1]{bertsekas2003convex} will be necessary and sufficient to characterize an optimal solution.

For technical reasons, we will further assume that the state-input pair takes values in some bounded set:

\begin{standing}\label{standing:bounded}
	$(x,u) \in \Omega$, where $\Omega \subset \R^n \times \R^m$ is a convex polytope.
\end{standing}

\section{A neural network-based representation of hybrid systems}\label{sec:neural_network_representation}
\begin{figure*}[t!]
	\centering
	\ifTwoColumn
		\includegraphics[width=.8\textwidth]{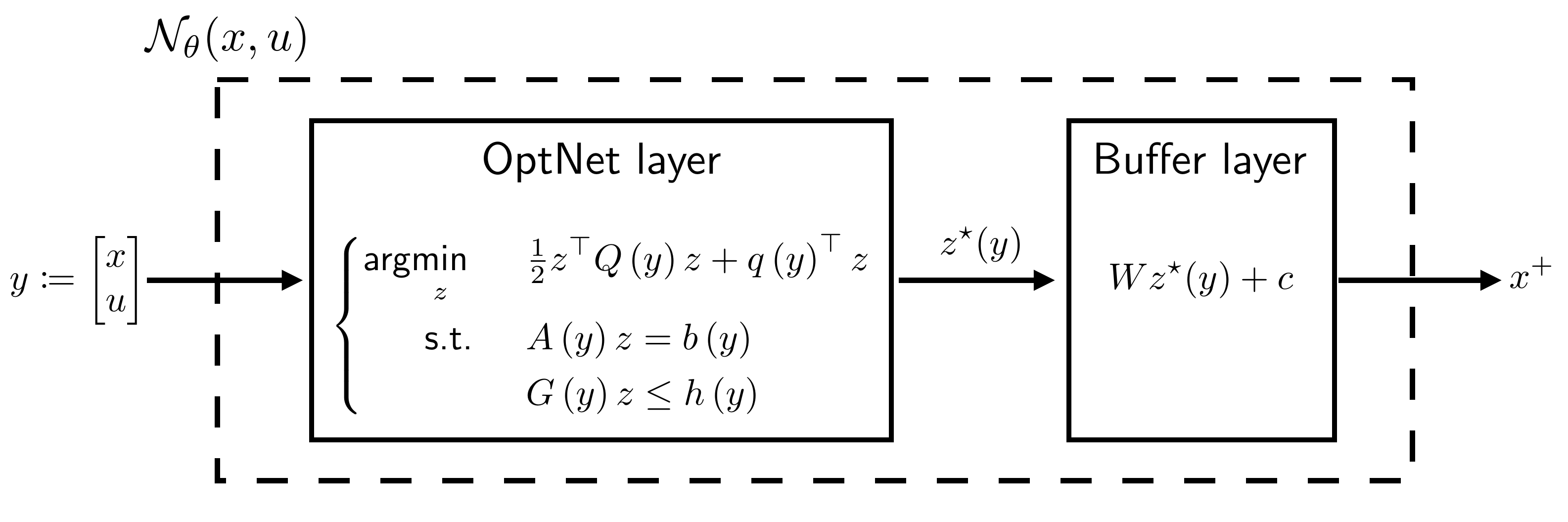}
	\else
		\includegraphics[width=\textwidth]{NN_architecture}
	\fi
	\caption{Neural network architecture considered in this work, which is composed by the cascade of an OptNet layer with a buffer one, which performs an affine transformation.}
	\label{fig:NN_architecture}
\end{figure*}

To address the problem introduced in \S \ref{sec:problem_description}, we will design $\mc N_\theta$ as a two-layer \gls{NN} with the simplified architecture illustrated in Fig.~\ref{fig:NN_architecture}, and then make use of available data $\{(x^{(i)},u^{(i)}, x^{+,{(i)}})\}_{i=1}^N$ to train the associated parameters, generically described by $\theta\in \R^p$. In particular, the underlying \gls{NN} consists of  an OptNet layer \cite{amos2017optnet,amos2018differentiable}, which takes the input pair $\col(x, u) \reqdef y$ as a parameter to solve the following generic \gls{QP}:
\begin{equation}\label{eq:optnet}
	\left.
	\begin{aligned}
		&\underset{z}{\textrm{min}} && \tfrac{1}{2} z^\top Q\left(y\right) z+q\left(y\right)^\top z \\
		&~\textrm{s.t. } && A\left(y\right) z=b\left(y\right), \\
		&&& G\left(y\right) z \leq h\left(y\right),
	\end{aligned}
	\right.
\end{equation}
thus returning an optimal solution $z^\star : \R^n \times \R^m \to \R^s$ as output.   This output is then passed through a linear layer with tailored matrix $W \in \R^{n \times s}$ and affine term $c \in \R^n$: 
\begin{equation}\label{eq:neural_network}
	x^+ = W z^\star(x, u) + c \reqdef \mc N_\theta(x,u).
\end{equation}
To ensure a unique optimizer for \eqref{eq:optnet} so that there is no ambiguity in the definition of the dynamics in \eqref{eq:neural_network}, during the training phase we will impose $Q\in\bbS^s_{\succ 0}$. This will be made possible in view of the main technical features possessed by the \gls{NN} in Fig.~\ref{fig:NN_architecture}, established next:


\begin{proposition}\label{prop:NN_properties}
	The {\gls{NN}} $\mc N_\theta : \R^p \times \R^n \times \R^m \to \R^n$ in \eqref{eq:neural_network} enjoys the following properties:
	\begin{itemize}
		\item[(i)] In \eqref{eq:optnet}, assume that $Q\in\bbS^s_{\succ 0}$ and that $A$ has full row rank. Then the output of $\mc N_\theta$ in \eqref{eq:neural_network} is differentiable \gls{wrt} the whole set of parameters $\theta$;
		\item[(ii)] $\mc N_\theta$ can represent any continuous {\gls{PWA}} mapping defined over $\Omega$. Specifically, in case the mapping to be modelled is defined over a regular partition of $\Omega$ with $r$ pieces, then the number of constraints that we require to reproduce it is no more than $2nr$, i.e., $l \le 2 n r$, and $s\le2n$.
	\end{itemize}
\end{proposition}
\begin{pf}
%
(i) Since the composition of differentiable mappings remains differentiable, the proof follows by combining \cite[Th.~1]{amos2017optnet}, which proves differentiability of $z^\star(x,u)$ \gls{wrt} the OptNet parameters under the postulated conditions on matrices $Q$ and $A$, and the fact that any successor state $x^+$ is defined by an affine combination of $z^\star(x, u)$ for any state-input pair $(x, u)$, which is hence differentiable \gls{wrt} both $W$ and $c$.\\
(ii) This part follows from Standing Assumption~\ref{standing:bounded} and \cite[Th.~2]{hempel2014inverse} directly. In fact, if one chooses $c=0$, and parameters $Q, q, A, b, G, h$ so that \eqref{eq:optnet} becomes:
\begin{equation}\label{eq:optnet_PWA}
	\left.
	\begin{aligned}
		&\underset{z}{\mathrm{min}} && \tfrac{1}{2} z^\top Q z+ (p + R y)^\top z\\
		&~\mathrm{s.t. } && F z + G y \leq h,
	\end{aligned}
	\right.
\end{equation}
with $Q\in\bbS^s_{\succ 0}$, then the structure in \cite[Eq.~4]{hempel2014inverse} is immediately recovered. Specifically, the inequality constraints in \eqref{eq:optnet_PWA} are obtained from those in \eqref{eq:optnet} by simply imposing $G(y)=F$ and $h(y)=h-G y$, for appropriate matrices $F$, $G$ and vector $h$ with row dimension $l\le2nr$ and $s\le2n$ columns (notice the slight abuse of notation). Remarkably, these choices are always possible since differentiability of the \gls{NN} output \gls{wrt} $\theta$, which follows from the first property shown in this statement, implies that $\mc N_\theta$ is an end-to-end trainable \gls{NN}.
\end{pf}

\sloppy In the rest of the paper, when we will refer to the OptNet layer we will tacitly consider the structure in \eqref{eq:optnet_PWA}. The set of parameters characterizing $\mc N_\theta$ is then $\theta \eqdef \{(Q, R, p, F, G, h), W, c\} \in \R^p$, which shall be determined during an offline training procedure using the available dataset $\{(x^{(i)},u^{(i)}, x^{+,{(i)}})\}_{i=1}^N$. With this regard, Proposition~\ref{prop:NN_properties}.(i) makes $\mc N_\theta$ in \eqref{eq:neural_network} an end-to-end trainable \gls{NN}, meaning that one is actually able to set values for the parameters $\theta$ exploiting available samples, as commonly happens with any other \gls{NN}. A differentiable output enables the use of a typical backpropagation strategy to find the gradient of the training loss function. Thus, one can train $\mc N_\theta$ with standard gradient descent methods for finding a minimum of that function.


As a main consequence of the second property in Proposition~\ref{prop:NN_properties}, instead, we note that \eqref{eq:OCP} turns out to coincide with an \gls{OCP} involving a continuous \gls{PWA} dynamics. 
The latter is known to be equivalent to a number of hybrid system model classes \cite{heemels2001equivalence}, such as \gls{LC} \cite{jin2022learning} or \gls{MLD} models \cite{bemporad1999control}. For this reason, the family of continuous \gls{PWA} dynamics excels in capturing a wide variety of real-world system behaviours. In particular, since $\mc N_\theta$ will be trained on the basis of available data $\{(x^{(i)},u^{(i)}, x^{+,{(i)}})\}_{i=1}^N$, \eqref{eq:OCP} actually amounts to an \gls{OCP} involving a data-based hybrid system representation.

Finally, note that the architecture expressed in \eqref{eq:neural_network}--\eqref{eq:optnet_PWA} represents a totally valid \gls{NN} in the machine learning realm, featuring attractive computational properties \cite{blondel2022efficient}. In particular, it incorporates an \emph{implicit layer} \cite{amos2017optnet,blondel2022efficient}, which requires an iterative process to compute its output, unlike traditional \emph{explicit layers}. Since the proposed method involves not only solving the \gls{QP} in \eqref{eq:optnet_PWA} but also learning its parameters, referring to it as a \gls{NN} is consistent with the terminology used in the literature.

\section{The \gls{OCP} in \eqref{eq:OCP} is a data-based \gls{MPCC}}\label{sec:OCP}
We derive next an equivalent formulation for the \gls{OCP} in \eqref{eq:OCP}.
Successively, we will then present in \S \ref{subsec:MPCCpriorart} available results that are applicable to \glspl{MPCC} models with the particular structure we obtain, before applying those results to our problem in \S\ref{sec:main_result}.

\subsection{Mathematical derivation}
Suppose that we have trained $\mc N_\theta$ in \eqref{eq:neural_network} to obtain the structure of the OptNet layer as in \eqref{eq:optnet_PWA} with $Q\in\bbS^s_{\succ 0}$, and with $c=0$.   The following \gls{KKT} system is then necessary and sufficient to characterize the optimal solution $z^\star$:
\begin{equation}\label{eq:KKT}
	\left\{
	\begin{aligned}
		&Q z^\star +R y+p+F^{\top} \lambda =0,\\
		&0 \leq (h-F z^\star-G y) \perp \lambda \geq 0,
	\end{aligned}
	\right.
\end{equation}
where $\lambda \in \R^l_{\ge0}$ represents the vector of Lagrange multipliers associated to the linear constraints. By recalling that $y = \col(x,u)$, from the \gls{KKT} system above we obtain a so-called \gls{LC} model of the system dynamics \cite{heemels2003complementarity} associated with the \gls{NN} $\mc N_\theta$, with architecture as in Fig.~\ref{fig:NN_architecture}:
\begin{equation}\label{eq:LC_model}
	\ifTwoColumn
		\left\{
		\begin{aligned}
			&x^{+} =-W Q^{-1} R\begin{bmatrix}
				x \\
				u
			\end{bmatrix}-W Q^{-1} F^{\top} \lambda-W Q^{-1} p, \\
			&0  \leq \left(F Q^{-1} R-G\right)\begin{bmatrix}
				x \\
				u
			\end{bmatrix}+F Q^{-1} F^{\top} \lambda\\
		&\hspace{4.2cm}+F Q^{-1} p+h \perp \lambda \geq 0.
		\end{aligned}
		\right.
	\else
		\left\{
		\begin{aligned}
			&x^{+} =-W Q^{-1} R\begin{bmatrix}
				x \\
				u
			\end{bmatrix}-W Q^{-1} F^{\top} \lambda-W Q^{-1} p, \\
			&0  \leq \left(F Q^{-1} R-G\right)\begin{bmatrix}
				x \\
				u
			\end{bmatrix}+F Q^{-1} F^{\top} \lambda+F Q^{-1} p+h \perp \lambda \geq 0.
		\end{aligned}
		\right.
	\fi
\end{equation}

Note that all of the terms in \eqref{eq:LC_model}  except $(x,u)$ and $\lambda$ are determined through a training procedure for $\mc N_\theta$ in \eqref{eq:neural_network}. Hence the terminology ``data-based representation''. 

Substituting our \gls{LC} model from \eqref{eq:LC_model} into \eqref{eq:OCP} yields:
\begin{equation}\label{eq:OCP_LC}
	\ifTwoColumn
		\left.
		\begin{aligned}
			&\underset{\bs{u}, \bs{x}, \bs{w}}{\textrm{min}} && \ell_T\left(x_T\right)+\sum_{t\in\mc T} \ell_t\left(x_t, u_t\right)\\
			& ~\textrm{ s.t. } && x_{t+1} = A x_t + B_u u_t + B_w w_t + d, \ \forall t \in \mc T,\\
			&&& 0 \!\le\! E_w w_t + E_x x_t  +E_u u_t  + e \perp w_t \!\geq\! 0, \ \forall t \in \mc T,\\
			&&& x_0 = x(0),
		\end{aligned}
		\right.
	\else
		\left.
		\begin{aligned}
			&\underset{\bs{u}, \bs{x}, \bs{w}}{\textrm{min}} && \ell_T\left(x_T\right)+\sum_{t\in\mc T} \ell_t\left(x_t, u_t\right)\\
			& ~\textrm{ s.t. } && x_{t+1} = A x_t + B_u u_t + B_w w_t + d, \text{ for all } t \in \mc T,\\
			&&& 0 \le E_w w_t + E_x x_t  +E_u u_t  + e \perp w_t \geq 0, \text{ for all } t \in \mc T,\\
			&&& x_0 = x(0),
		\end{aligned}
		\right.	
	\fi
\end{equation}
where the matrices $A$, $B_u$, $B_w$, $E_w$, $E_x$, $E_u$, and vectors $d$ and $e$ can be obtained by rearranging the terms and definition of the complementarity variable $w \eqdef \lambda$. 
Specifically, $B_w \eqdef -W Q^{-1} F^{\top}$, $d \eqdef -W Q^{-1} p$, $E_w \eqdef F Q^{-1} F^{\top}$, $e \eqdef F Q^{-1} p+h$.  The matrices $A$ and $B_u$ (respectively, $E_x$ and $E_u$) follow by partitioning $-W Q^{-1} R$ (resp., $F Q^{-1} R-G$) with $n+m$ columns into two matrices with $n$ and $m$ columns, respectively. The notation is then analogous to \eqref{eq:OCP}, and $\bs w \eqdef \col((w_k)_{t \in \mc T})$ denotes the trajectory of complementarity variables corresponding to $\bs u$ and $\bs x$. 
A distinct feature of the \gls{LC} dynamics are the complementarity constraints inherited from \eqref{eq:LC_model}, which hence turns the \gls{OCP} \eqref{eq:OCP_LC}, equivalent to \eqref{eq:OCP} under our choice of $\mc N_\theta$ in \eqref{eq:neural_network}, into a data-based \gls{MPCC}. 

\begin{remark}
	Unlike \cite{jin2022learning}, where an {\gls{LC}} model shall be inferred from data directly, we obtain it as a consequence of choosing to train a {\gls{NN}} $\mc N_\theta$ with architecture as in Fig.~\ref{fig:NN_architecture}. For this reason, we do not require strict complementarity of the resulting {\gls{LC}} model to recover differentiability \gls{wrt} the main parameters, as postulated instead in \cite{jin2022learning,de2018end}. 
\end{remark}

\subsection{Prior results on a class of \glspl{MPCC}}\label{subsec:MPCCpriorart}
In general, \glspl{MPCC} amount to nonlinear, nonconvex optimization problems that can be very challenging to solve \cite{ralph2008mathematical}. Indeed, for such problems the standard constraint qualifications for \gls{NLP} (e.g., the classical Mangasarian-Fromovitz one \cite[\S 5.5.4]{bertsekas2003convex}) typically fail to hold at any feasible point \cite{chen1995nonlinear}. 

On the other hand, it has been recently proven in \cite{hempel2017strong} that if the \gls{MPCC} has a specific structure then one is able to establish strong stationarity conditions characterizing a (local) optimal solution. In particular, by referring to the data-based \gls{MPCC} in \eqref{eq:OCP_LC} the following three requirements on the \gls{LC} model are sufficient to recover the strong stationarity conditions derived in \cite[Th.~1]{hempel2017strong}:
\begin{condition}\label{ass:deterministic_behaviour}
	Given any state-input pair $(x, u)$, every complementarity variable $w$ that solves $0 \le E_w w + E_x x  +E_u u  + e \perp w \geq 0$ results in the same successor state $x^{+}$.
	
\end{condition}
\begin{condition}\label{ass:decomposition}
	The complementarity problem $0 \le E_w w + E_x x  +E_u u  + e \perp w \geq 0$ can be decomposed elementwise \gls{wrt} the complementarity variable $w$, i.e.,
	\begin{equation}\label{eq:decomposition}
		\ifTwoColumn
			\forall i \in \{1, \ldots, l\}: 0 \leq m_i w_i + D_i x + G_i u + e_i \perp w_i \geq 0,
		\else
			\forall i \in \{1, \ldots, l\}: \quad 0 \leq m_i w_i + D_i x + G_i u + e_i \perp w_i \geq 0,
		\fi
	\end{equation}
	for $m_i > 0$ so that $E_w = \diag((m_i)_{i=1}^l)$, $E_x = \normaltext{\col}((D_i)_{i=1}^l)$, $E_u = \normaltext{\col}((G_i)_{i=1}^l)$ and $e=\normaltext{\col}((e_i)_{i=1}^l)$.
	
\end{condition}
\begin{condition}\label{ass:nontrivial}
	Given any state-input pair $(x, u)$, there exists some $i \in \{1, \ldots, l\}$ such that $D_i x+G_i u+e_i<0$.
	
\end{condition}

While Condition~\ref{ass:deterministic_behaviour} guarantees the well-posedness of the \gls{LC} model we consider, entailing a deterministic behaviour for the resulting dynamics, Conditions~\ref{ass:decomposition} and \ref{ass:nontrivial} are rather technical and partially limit the \gls{LC} models we are allowed to consider. Specifically, Condition~\ref{ass:decomposition} means that the solution set of the complementarity problem $0 \le E_w w + E_x x  +E_u u  + e \perp w \geq 0$ is given by the cartesian product of the solution sets of \eqref{eq:decomposition}, while Condition~\ref{ass:nontrivial} requires the existence of a solution $w_i \neq 0$ to \eqref{eq:decomposition}, for any fixed pair $(x, u)$.
\begin{remark}
	We consider an elementwise decomposition of the complementarity problem $0 \le E_w w + E_x x  +E_u u  + e \perp w \geq 0$ for simplicity, although a generalization of Condition~\ref{ass:decomposition} allowing for a block-diagonal decomposition is also possible -- see {\cite[Ass.~2]{hempel2017strong}}.
\end{remark}

Armed with these requirements, our next result provides necessary and sufficient conditions to characterize a local solution to the optimal control \gls{MPCC} in \eqref{eq:OCP_LC}:
\begin{lemma}\textup{(\hspace{-.03em}\cite[Th.~1]{hempel2017strong})}\label{lemma:strong_stationarity}
	Let $\left(\bs x^\star, \bs u^\star, \bs w^\star\right)$ be feasible for the {\gls{MPCC}} in \eqref{eq:OCP_LC} with an {\gls{LC}} model satisfying Conditions~\ref{ass:deterministic_behaviour}--\ref{ass:nontrivial}. Then, $\left(\bs x^\star, \bs u^\star, \bs w^\star\right)$ is locally optimal if and only if the standard {\gls{KKT}} conditions for \eqref{eq:OCP_LC} admit a primal-dual solution pair.
\end{lemma}

The statement in Lemma~\ref{lemma:strong_stationarity} enables one to seek a local solution to the optimal control \gls{MPCC} \eqref{eq:OCP_LC} through the solution of a classical \gls{KKT} system, i.e., as an \gls{NLP}. This is typically computationally advantageous with significantly better scaling features compared to more traditional mixed-integer approaches to solving \glspl{OCP} based on generic \gls{PWA} or hybrid models%
\footnote{
	For NP-hard problems as \eqref{eq:OCP_LC} one should not expect globally optimal solutions to be efficiently computable in general \cite{daafouz2009switched}, regardless of the method employed.
}
\cite{hempel2017strong,hall2021sequential}. 

Then, if we manage to train our \gls{NN} $\mc N_\theta$ in \eqref{eq:neural_network} so that Conditions~\ref{ass:deterministic_behaviour}--\ref{ass:nontrivial} are satisfied, $\mc N_\theta$ would also bring
 major computational advantages when used as part of an \gls{OCP}.
\section{Main results}\label{sec:main_result}
We now show how the \gls{NN} $\mc N_\theta$ in \eqref{eq:neural_network}, where $z^\star(x,u)$ minimizes \eqref{eq:optnet} (or \eqref{eq:optnet_PWA}) for a given pair $(x,u) \in \Omega$, can be trained so that Conditions~\ref{ass:deterministic_behaviour}--\ref{ass:nontrivial} are met, thereby enabling us to solve the data-based \gls{OCP} \eqref{eq:OCP_LC} as an \gls{NLP}.

Specifically, in \cite{hempel2017strong} it was shown that the following inverse optimization model of \gls{PWA} dynamics leads to an \gls{LC} model so that, once it is embedded into an \gls{OCP} as  in \eqref{eq:OCP}, we can satisfy all three conditions collectively:
\ifTwoColumn
	\begin{subequations}\label{eq:alfa_beta}
		\begin{align}
			& x^{+}=\alpha^\star(x,u)-\beta^\star(x,u)=\alpha^\star-\beta^\star, \text{ with } \label{eq:first}\\
			&\notag\\
			& \alpha^\star = \left\{
			\begin{aligned}
				&\underset{\alpha \in \mathbb{R}^n}{\textrm{argmin}} && \tfrac{1}{2}\|\alpha-A_\psi x - B_\psi u- c_\psi\|_{Q_\alpha}^2 \\
				&~~~~\textrm{ s.t. } && \alpha \geq A_{\alpha, i} x+ B_{\alpha, i} u +c_{\alpha, i},\\
				&&&\hspace{2.5cm}\textrm{ for all } i \in \{1,\ldots, n_r^\alpha\}
			\end{aligned}
			\right.\label{eq:second}\\
			& \beta^\star = \left\{
			\begin{aligned}
				&\underset{\beta \in \mathbb{R}^n}{\textrm{argmin}} && \tfrac{1}{2}\|\beta-A_\phi x - B_\phi u - c_\phi\|_{Q_\beta}^2 \\
				&~~~~\textrm{ s.t. } && \beta \geq A_{\beta, j} x + B_{\beta, j} u + c_{\beta, j},\\
				&&&\hspace{2.4cm}\textrm{ for all } j \in \{1,\ldots,n_r^\beta\},
			\end{aligned}
			\right.\label{eq:third}
		\end{align}
	\end{subequations}
\else
	\begin{subequations}\label{eq:alfa_beta}
		\begin{align}
			&  x^{+}=\alpha^\star(x,u)-\beta^\star(x,u)=\alpha^\star-\beta^\star, \text{ with } \label{eq:first}\\
			&\notag\\
			& \alpha^\star = \left\{
			\begin{aligned}
				&\underset{\alpha \in \mathbb{R}^n}{\textrm{argmin}} && \tfrac{1}{2}\|\alpha-A_\psi x - B_\psi u- c_\psi\|_{Q_\alpha}^2 \\
				&~~~~\textrm{ s.t. } && \alpha \geq A_{\alpha, i} x+ B_{\alpha, i} u +c_{\alpha, i}, \textrm{ for all } i \in \{1,\ldots, n_r^\alpha\}
			\end{aligned}
			\right.\label{eq:second}\\
			& \beta^\star = \left\{
			\begin{aligned}
				&\underset{\beta \in \mathbb{R}^n}{\textrm{argmin}} && \tfrac{1}{2}\|\beta-A_\phi x - B_\phi u - c_\phi\|_{Q_\beta}^2 \\
				&~~~~\textrm{ s.t. } && \beta \geq A_{\beta, j} x + B_{\beta, j} u + c_{\beta, j}, \textrm{ for all } j \in \{1,\ldots,n_r^\beta\},
			\end{aligned}
			\right.\label{eq:third}
		\end{align}
	\end{subequations}
\fi
\sloppy where $Q_\alpha$, $Q_\beta \in \mathbb{S}^n_{\succ0}$ are diagonal, but otherwise arbitrary, matrices. Moreover, the elements $\{(A_{\alpha, i},B_{\alpha, i},c_{\alpha, i})\}_{i=1}^{n_r^\alpha}$ and $\{(A_{\beta, j},B_{\beta, j},c_{\beta, j})\}_{j=1}^{n_r^\beta}$ were determined in \cite{hempel2017strong} on the basis of a \gls{PWA} partitions, computed through available methods as, e.g., Delaunay triangulations and Voronoi diagrams, of the state-input space $\Omega \subset \R^n\times\R^m$ introduced in Standing Assumption~\ref{standing:bounded} -- see, for instance, the discussion in \cite{hempel2014inverse,hempel2013every}. 
%
In addition, the affine functions in the costs originating from $(A_\psi,B_\psi,c_\psi)$ (respectively, $(A_\phi,B_\phi,c_\phi)$), instead, were required to lower bound $\{(A_{\alpha, i},B_{\alpha, i},c_{\alpha, i})\}_{i=1}^{n_r^\alpha}$ (resp., $\{(A_{\beta, j},B_{\beta, j},c_{\beta, j})\}_{j=1}^{n_r^\beta}$) in each region of the underlying partition. 
%
Specifically, one has to satisfy the following:
\begin{equation}\label{eq:strict_inequalities}
	\ifTwoColumn
		\begin{aligned}
			&A_\psi x + B_\psi u + c_\psi\\
			&~~~~~~~~~<\underset{i \in \{1,\ldots,n_r^\alpha\}}{\textrm{max}}~ \{A_{\alpha, i} x+B_{\alpha, i} u+c_{\alpha, i}\}, \ \forall (x, u) \in \Omega,\\
			&A_\phi x + B_\phi u + c_\phi\\
			&~~~~~~~~~<\underset{j \in \{1,\ldots,n_r^\beta\}}{\textrm{max}}~ \{A_{\beta, j} x+B_{\beta, j} u+c_{\beta, j}\}, \ \forall (x, u) \in \Omega.
		\end{aligned}	
	\else
		\begin{aligned}
			&A_\psi x + B_\psi u + c_\psi<\underset{i \in \{1,\ldots,n_r^\alpha\}}{\textrm{max}}~ \{A_{\alpha, i} x+B_{\alpha, i} u+c_{\alpha, i}\}, \text{ for all } (x, u) \in \Omega,\\
			&A_\phi x + B_\phi u + c_\phi<\underset{j \in \{1,\ldots,n_r^\beta\}}{\textrm{max}}~ \{A_{\beta, j} x+B_{\beta, j} u+c_{\beta, j}\}, \text{ for all } (x, u) \in \Omega.
		\end{aligned}
	\fi
\end{equation}
%
%
%
As suggested in \cite{hempel2017strong}, note that the above can always be satisfied for a given \gls{PWA} system, e.g., by choosing any index $i \in \{1,\ldots,n_r^\alpha\}$ and $j \in \{1,\ldots,n_r^\beta\}$ and by setting
\begin{equation}\label{eq:possible_choice}
	\begin{aligned}
		& A_\psi x + B_\psi u + c_\psi=A_{\alpha, i} x+B_{\alpha, i} u+c_{\alpha, i}-\eta, \\
		& A_\phi x + B_\phi u + c_\phi=A_{\beta, j} x+B_{\beta, j} u+c_{\beta, j}-\zeta,
	\end{aligned}
\end{equation}
with arbitrary terms $\eta$, $\zeta > 0$. 

\begin{figure*}[t!]
	\centering
	\ifTwoColumn
	\includegraphics[width=.8\textwidth,trim=3cm 15cm 3cm 0cm,clip]{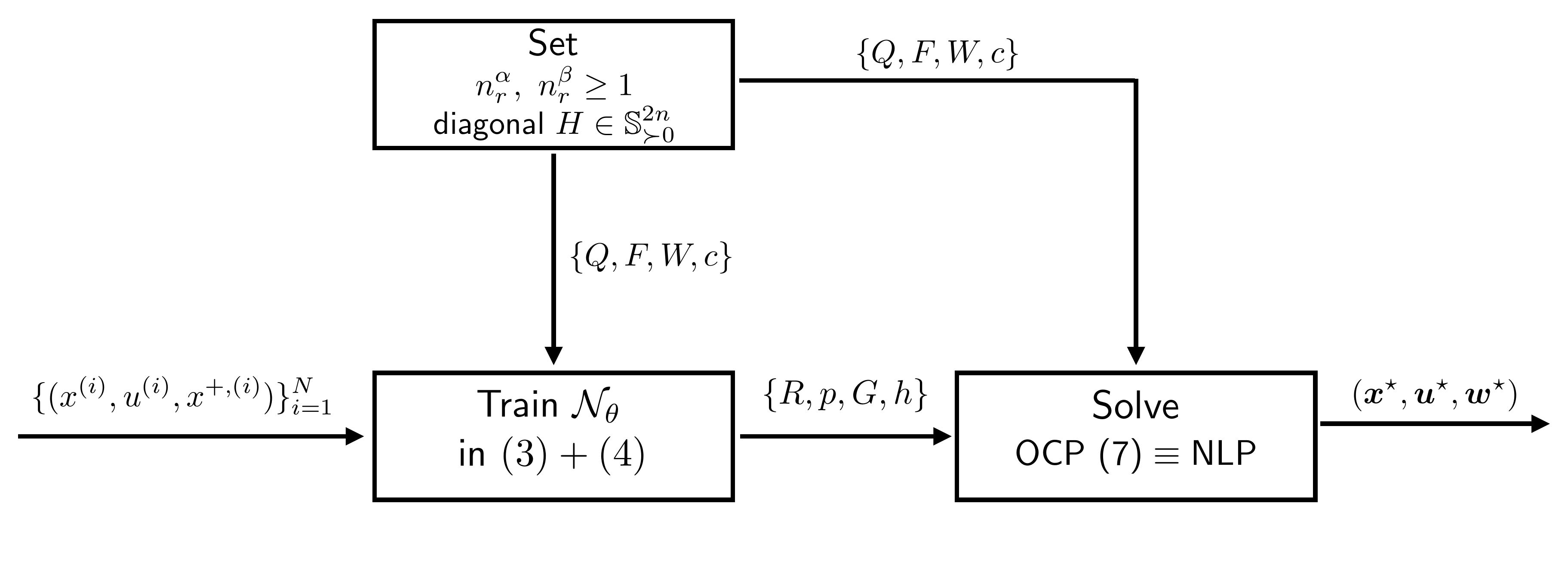}
	\else
	\includegraphics[width=\textwidth,trim=3cm 15cm 3cm 0cm,clip]{data_workflow}
	\fi
	\caption{Proposed \gls{NN}-based hybrid system identification for control approach. After choosing hyperparameters $n_r^\alpha$, $n_r^\beta\ge1$ and diagonal $H\in\bbS^{2n}_{\succ 0}$, one directly obtains elements $Q$, $F$, $W$, and $c$ to be used, along with the available data $\{(x^{(i)},u^{(i)}, x^{+,{(i)}})\}_{i=1}^N$, to train the \gls{NN} $\mc N_\theta$ in \eqref{eq:neural_network} and \eqref{eq:optnet_PWA} -- see Theorem~\ref{th:nn_representation}. Plugging all the elements in \eqref{eq:OCP_LC} then enables to find some locally optimal triplet $\left(\bs x^\star, \bs u^\star, \bs w^\star\right)$ by solving the \gls{KKT} system associated to the underlying \gls{OCP} problem. 
	}
	\label{fig:data_workflow}
\end{figure*}

By taking the \gls{NN}-based perspective of in this work, however, one does not need to compute all of the aforementioned elements explicitly. As will be made clear in our main result, once the number of regions determining each partition is fixed, i.e., $n_r^\alpha$, $n_r^\beta\ge1$ (which will hence coincide with the hyperparameters of our data-driven approach, along with $Q_\alpha$, $Q_\beta$), we will obtain matrices and vectors $\{(A_{\alpha, i},B_{\alpha, i},c_{\alpha, i})\}_{i=1}^{n_r^\alpha}$, $\{(A_{\beta, j},B_{\beta, j},c_{\beta, j})\}_{j=1}^{n_r^\beta}$, $(A_\psi,B_\psi,c_\psi)$ and $(A_\phi,B_\phi,c_\phi)$, as weights of $\mc N_\theta$ in \eqref{eq:neural_network}.  Specifically, we can claim the following:
\begin{theorem}\label{th:nn_representation}
	Let $n_r^\alpha$, $n_r^\beta\ge1$ be fixed. The {\gls{NN}} $\mc N_\theta : \R^p \times \R^n \times \R^m \to \R^n$ in \eqref{eq:neural_network} can be trained to produce a hybrid system model satisfying Conditions~\ref{ass:deterministic_behaviour}--\ref{ass:nontrivial}.
\end{theorem}
\begin{pf}
	To start we note that, for given $n_r^\alpha$, $n_r^\beta\ge1$ and any state-input pair $(x,u) \in \Omega$, \eqref{eq:second} and \eqref{eq:third} can be lumped together to obtain the following separable \gls{QP}:
	\ifTwoColumn
		$$
		\left.
		\begin{aligned}
			&\underset{(\alpha,\beta)\in \R^{2n}}{\textrm{min}} && \frac{1}{2} \left\|\begin{bmatrix} \alpha\\\beta\end{bmatrix} - \begin{bmatrix}A_\psi\\A_\phi\end{bmatrix} x - \begin{bmatrix}B_\psi\\B_\phi\end{bmatrix} u - \begin{bmatrix}c_\psi\\c_\phi\end{bmatrix} \right\|_H^2\\
			&~~~~\textrm{ s.t. } && \begin{bmatrix}
				I_n\otimes\bsone_{n_r^\alpha} & \bs0_{n n_r^\alpha\times n}\\
				\bs0_{n n_r^\beta\times n} & I_n\otimes\bsone_{n_r^\beta}
			\end{bmatrix} \begin{bmatrix} \alpha\\\beta \end{bmatrix}\\
		&&&~~~~~~~~\ge \begin{bmatrix}
				A_{\alpha,1}\\
				\vdots\\
				A_{\alpha,n_r^\alpha}\\
				A_{\beta,1}\\
				\vdots\\
				A_{\beta,n_r^\beta}\\
			\end{bmatrix} x + \begin{bmatrix}
				B_{\alpha,1}\\
				\vdots\\
				B_{\alpha,n_r^\alpha}\\
				B_{\beta,1}\\
				\vdots\\
				B_{\beta,n_r^\beta}\\
			\end{bmatrix} u+ \begin{bmatrix}
				c_{\alpha,1}\\
				\vdots\\
				c_{\alpha,n_r^\alpha}\\
				c_{\beta,1}\\
				\vdots\\
				c_{\beta,n_r^\beta}\\
			\end{bmatrix},
		\end{aligned}
		\right.
		$$	
	\else
		$$
		\left.
		\begin{aligned}
			&\underset{(\alpha,\beta)\in \R^{2n}}{\textrm{min}} && \frac{1}{2} \left\|\begin{bmatrix} \alpha\\\beta\end{bmatrix} - \begin{bmatrix}A_\psi\\A_\phi\end{bmatrix} x - \begin{bmatrix}B_\psi\\B_\phi\end{bmatrix} u - \begin{bmatrix}c_\psi\\c_\phi\end{bmatrix} \right\|_H^2\\
			&~~~~\textrm{ s.t. } && \begin{bmatrix}
				I_n\otimes\bsone_{n_r^\alpha} & \bs0_{n n_r^\alpha\times n}\\
				\bs0_{n n_r^\beta\times n} & I_n\otimes\bsone_{n_r^\beta}
			\end{bmatrix} \begin{bmatrix} \alpha\\\beta \end{bmatrix} \ge \begin{bmatrix}
				A_{\alpha,1}\\
				\vdots\\
				A_{\alpha,n_r^\alpha}\\
				A_{\beta,1}\\
				\vdots\\
				A_{\beta,n_r^\beta}\\
			\end{bmatrix} x + \begin{bmatrix}
				B_{\alpha,1}\\
				\vdots\\
				B_{\alpha,n_r^\alpha}\\
				B_{\beta,1}\\
				\vdots\\
				B_{\beta,n_r^\beta}\\
			\end{bmatrix} u+ \begin{bmatrix}
				c_{\alpha,1}\\
				\vdots\\
				c_{\alpha,n_r^\alpha}\\
				c_{\beta,1}\\
				\vdots\\
				c_{\beta,n_r^\beta}\\
			\end{bmatrix},
		\end{aligned}
		\right.
		$$
	\fi
	where, for simplicity, we have imposed $\diag(Q_\alpha,Q_\beta) = H\in \mathbb{S}^{2n}_{\succ0}$. Then, by introducing the decision variable $\xi \eqdef \col(\alpha,\beta)$ and after suitably redefining all matrices/vectors, the \gls{QP} above admits a compact form as:
	\ifTwoColumn
	\begin{equation}\label{eq:single_QP_compact}
	\begin{aligned}
	&\left.
	\begin{aligned}
		&\underset{\xi\in \R^{2n}}{\textrm{min}} && \tfrac{1}{2} \| \xi - A_\gamma x - B_\gamma u - c_\gamma \|_H^2\\
		&~\textrm{ s.t. } && S_\xi \xi \ge A_\xi x + B_\xi u + c_\xi.
	\end{aligned}
	\right.\\
	\\
	&\implies\left.
		\begin{aligned}
			&\underset{\xi\in \R^{2n}}{\textrm{min}} && \tfrac{1}{2} \xi^\top H\xi - (A_\gamma x + B_\gamma u + c_\gamma)^\top H \xi\\
			&~\textrm{ s.t. } && S_\xi \xi \ge A_\xi x + B_\xi u + c_\xi.
		\end{aligned}
		\right.
	\end{aligned}
	\end{equation}
	\else
	\begin{equation}\label{eq:single_QP_compact}
		\left.
		\begin{aligned}
			&\underset{\xi\in \R^{2n}}{\textrm{min}} && \tfrac{1}{2} \| \xi - A_\gamma x - B_\gamma u - c_\gamma \|_H^2\\
			&~\textrm{ s.t. } && S_\xi \xi \ge A_\xi x + B_\xi u + c_\xi.
		\end{aligned}
		\right. \implies \left.
		\begin{aligned}
			&\underset{\xi\in \R^{2n}}{\textrm{min}} && \tfrac{1}{2} \xi^\top H\xi - (A_\gamma x + B_\gamma u + c_\gamma)^\top H \xi\\
			&~\textrm{ s.t. } && S_\xi \xi \ge A_\xi x + B_\xi u + c_\xi.
		\end{aligned}
		\right.
	\end{equation}
	\fi
	
	In particular, $A_\gamma \in \R^{2n \times n}$, $B_\gamma \in \R^{2n \times m}$, $c_\gamma \in \R^{2n}$, while $A_\xi \in \R^{n(n_r^\alpha+n_r^\beta) \times n}$, $B_\xi \in \R^{n(n_r^\alpha+n_r^\beta) \times m}$ and $c_\xi \in \R^{n(n_r^\alpha+n_r^\beta)}$. Note that, in addition, the matrix $S_\xi \in \R^{n(n_r^\alpha+n_r^\beta) \times 2n}$ is a full column rank matrix for any $n_r^\alpha$, $n_r^\beta \ge 1$. The statement then follows by making a one-to-one correspondence between the model in \eqref{eq:neural_network}, where $z^\star(x,u)$ minimizes \eqref{eq:optnet_PWA} for some pair $(x,u) \in \Omega$ with affine layer in cascade, and \eqref{eq:single_QP_compact}. In particular, for given $n_r^\alpha$, $n_r^\beta\ge1$, and diagonal $H\in \mathbb{S}^{2n}_{\succ0}$, to fully recover \eqref{eq:single_QP_compact} while training $\mc N_\theta$ one can treat $z$ in \eqref{eq:optnet_PWA} as a decision variable living in $\R^{2n}$, additionally imposing $Q=H \in \mathbb{S}^{2n}_{\succ0}$, $F=-S_\xi$, $W=[I_n \; -I_n]$, and $c=0$. Thus, once trained the \gls{NN} $\mc N_\theta$, we obtain equivalences: $-H^{-1} R=[A_\gamma \; B_\gamma]$, $-H^{-1} p=c_\gamma$, $G = [A_\xi \; B_\xi]$, and $-h=c_\xi$.
	To conclude, note that the required conditions on parameters $\theta$, specifically, on the fixed matrices/vector $Q$, $F$, $W$, and $c$ can be easily imposed during the training as a direct consequence of Proposition~\ref{prop:NN_properties}.(i).	
\end{pf}

Under a careful choice of some of the weights characterizing the \gls{NN} $\mc N_\theta$ in \eqref{eq:neural_network}, the latter can then be trained by means of samples $\{(x^{(i)},u^{(i)}, x^{+,{(i)}})\}_{i=1}^N$ so that a data-based \gls{LC} model meeting Conditions~\ref{ass:deterministic_behaviour}--\ref{ass:nontrivial} is actually produced. In particular, our data-driven approach reduces the number of parameters that must be set during the training phase of $\mc N_\theta$ to $\{R, p, G, h\}$ -- see Fig.~\ref{fig:data_workflow} for a schematic representation. Alternatively, one may also decide to train the structure in \eqref{eq:alfa_beta} directly, and then successively recover the weights in \eqref{eq:neural_network}--\eqref{eq:optnet_PWA}. Specifically, learning the model in \eqref{eq:alfa_beta} amounts to determine the elements $\{(A_{\alpha, i},B_{\alpha, i},c_{\alpha, i})\}_{i=1}^{n_r^\alpha}$, $\{(A_{\beta, j},B_{\beta, j},c_{\beta, j})\}_{j=1}^{n_r^\beta}$, $(A_\psi,B_\psi,c_\psi)$, and  $(A_\phi,B_\phi,c_\phi)$, with hyperparameters $n_r^\alpha$ and $n_r^\beta$. To this end, note that also the choice of the Hessian matrix in \eqref{eq:second} and \eqref{eq:third}, i.e., $Q_\alpha$ and $Q_\beta$, is arbitrary.

To conclude, Corollary~\ref{cor:opt_control_nn} essentially particularizes the result in Lemma~\ref{lemma:strong_stationarity} to the optimal control of the data-based hybrid model designed through the \gls{NN} $\mc N_\theta$ in Fig.~\ref{fig:NN_architecture}:

\begin{corollary}\label{cor:opt_control_nn}
	Let the {\gls{NN}} $\mc N_\theta : \R^p \times \R^n \times \R^m \to \R^n$ in \eqref{eq:neural_network} be trained to produce a hybrid system model satisfying Conditions~\ref{ass:deterministic_behaviour}--\ref{ass:nontrivial}, along with the inequalities in \eqref{eq:strict_inequalities}. 
	 Let $\left(\bs x^\star, \bs u^\star, \bs w^\star\right)$ be feasible for the resulting {\gls{OCP}} in the form of \eqref{eq:OCP_LC}. Then, $\left(\bs x^\star, \bs u^\star, \bs w^\star\right)$ is locally optimal if and only if the standard {\gls{KKT}} conditions for \eqref{eq:OCP_LC} admit a primal-dual solution pair.
\end{corollary}

%

Corollary~\ref{cor:opt_control_nn} requires that if our \gls{NN} model $\mc N_\theta$ is to be included in the \gls{OCP} \eqref{eq:OCP_LC}, its parameters must also satisfy the strict inequalities in \eqref{eq:strict_inequalities} in order for the local optimality results of Lemma~\ref{lemma:strong_stationarity} to apply. 
However, as common to any type of \gls{NN}, these inequalities can not be enforced uniformly over $\Omega$ during the training phase, i.e., strict inequalities are guaranteed to hold true for the considered (state-input)-successor state samples only, $\{(x^{(i)},u^{(i)}, x^{+,{(i)}})\}_{i=1}^N$, and not for all $(x,u) \in \Omega$.  Therefore, some a-posteriori verification procedure shall be applied to make sure they are satisfied for all $(x,u) \in \Omega$. Conversely, the relations in \eqref{eq:possible_choice} provide us with a possible means of resolution by setting $(A_\psi,B_\psi,c_\psi)$ and $(A_\phi,B_\phi,c_\phi)$ to meet \eqref{eq:strict_inequalities}. In particular, imposing equalities in \eqref{eq:possible_choice} can be done either before or after the training process of $\mc N_\theta$. Also in this case, however, applying some a-posteriori verification procedure is inevitable.

\begin{remark}
	In contrast to \cite{yang2022modeling}, we give an explicit expression of a {\gls{NN}} $\mc N_\theta$ for which local stationarity conditions, coinciding with the standard {\gls{KKT}} system, are known to hold true for the {\gls{MPCC}} obtained once employed the {\gls{NN}} $\mc N_\theta$ as a data-based hybrid system model in the {\gls{OCP}} \eqref{eq:OCP}.
\end{remark}

\section{Numerical experiments}\label{sec:simulations}
We now verify the performance of $\mc N_\theta$ in \eqref{eq:neural_network}--\eqref{eq:optnet_PWA} to learning continuous \gls{PWA} models from dataset produced by \gls{PWA} and nonlinear dynamics. All simulations are run on a laptop equipped with a AMD R7 8840U processor.

To this end, we focus on the structure reported in \eqref{eq:alfa_beta} directly, since Theorem~\ref{th:nn_representation} establishes that the two representations are equivalent,
and we set $Q_\alpha=Q_\beta=I$, and $n_r^\alpha=n_r^\beta=7$ unless otherwise stated. 
We implement our method in Python, utilizing JAX~\cite{jax2018github}, JaxOPT~\cite{blondel2022efficient} and Scipy~\cite{2020SciPy}. The models have been trained to solve the following standard regularized training problem:
\begin{equation}\label{eq:training}
	\ifTwoColumn
	\left.
	\begin{aligned}
		&\underset{\theta}{\textrm{min}} & & \frac{1}{N} \sum_{i=1}^{N} \|x^{+,(i)}-\hat x^{+,(i)}\|_2^2 + \lambda \|\theta\|_2^2\\
		&\textrm{ s.t. } & & \hat x^{+,(i)}\!=\!\alpha_\theta(x^{(i)},u^{(i)})\!-\!\beta_\theta(x^{(i)},u^{(i)}),~\forall i=1,\dots,N,
	\end{aligned}	
	\right.
	\else
	\left.
	\begin{aligned}
		&\underset{\theta}{\textrm{min}} & & \frac{1}{N} \sum_{i=1}^{N} \|x^{+,(i)}-\hat x^{+,(i)}\|_2^2 + \lambda \|\theta\|_2^2\\
		&\textrm{ s.t. } & & \hat x^{+,(i)}=\alpha_\theta(x^{(i)},u^{(i)})-\beta_\theta(x^{(i)},u^{(i)}), \text{ for all }i=1,\dots,N,
	\end{aligned}	
	\right.
	\fi
\end{equation}
\sloppy where $\alpha_\theta(x,u)$ and $\beta_\theta(x,u)$ reminds the explicit dependence from $\theta = \{(A_{\alpha, i},B_{\alpha, i},c_{\alpha, i})\}_{i=1}^{7} \cup \{(A_{\beta, j},B_{\beta, j},c_{\beta, j})\}_{j=1}^{7} \cup \{(A_\psi,B_\psi,c_\psi), (A_\phi,B_\phi,c_\phi)\}$ of \eqref{eq:second} and \eqref{eq:third}, respectively.
The training problem was solved using the \texttt{SLSQP} algorithm \cite{kraft1988software} with $\lambda=0.01$, and $\mathtt{BoxCDQP}$ from JaxOPT as internal \gls{QP} solver. 

We then evaluate the approximation quality of the models obtained on the basis of open-loop predictions, produced by exciting the trained $\mc N_\theta$ with an unseen input sequence for 1000 time steps, and employing standard metrics \cite{breschi2016identification,breschi2016piecewise,mejari2018towards,masti2021learning} such as \gls{BFR}, defined as
$
	\mathrm{BFR}=\|\hat X^+ -X^+\|_F/\| X^+ - \bar X^+\|_F,
$
and the \gls{RMS} error,
$
	\mathrm{RMS}=\sqrt{\frac{1}{N} \sum_{i=1}^{N}\|x^{+,(i)}-\hat x^{+,(i)}\|_2^2}~.
$
Referring to \gls{BFR}, all matrices have dimension $n\times N$, with $X^+$ stacking the true data, $\bar X^+$ is the componentwise average vector of $x$, while $\hat X^+$ is the predicted value of $\hat X^+$ obtained in open-loop prediction. For system identification purposes, a high (respectively, low) value for \gls{BFR} (\gls{RMS}) is desirable.

Unless otherwise stated, all tests have been conducted by using 5000 normalized samples, adding white noise $N(0,0.01)$ to the sequence of state measurements.

\subsection{Comparison with PARC}\label{sub:parc}
We first contrast the performance achieved by our method with that obtained by \gls{PARC}, a state-of-the-art \gls{PWA} regression method that has been configured to perform up to 25 iterations and consider 20 clusters. 
To take into account the nonconvex nature of both learning problems, each test has been repeated for 10 different initial conditions of the optimizer.

\subsubsection{Piecewise system benchmarks}
To compare the two approaches on learning hybrid systems, we consider the following benchmarks:
\begin{itemize}
	\item A system described by the following equation:
	$$
		\Sigma_{\mathrm{PWA}}: x_{k+1}= A x_k + B u_k + W_A~\mathtt{clip}_{[0,2]}(W_B x_k),
	$$
	where $A$, $W_A$, $W_B \in \mathrm{R}^{4\times4}$, $B \in \mathrm{R}^{4\times2}$, and $\mathtt{clip}(\cdot)$ is a mapping that rounds the value of its argument in the range defined by the subscript. The entries of $A$, $B$ have been drawn from uniform distributions $U(0,1)$ and $U(0,1/3)$, respectively, while the entries of $W_A$, $W_B$ follow a normal distribution $N(0,1/2)$;
	
	\item The problem adopted in \cite[\S IV.A]{breschi2016identification}, which we simulate from a random initial condition using an excitation signal with the features described in \cite{breschi2016identification}.
 We will refer to this system, which has also been used as benchmark in~\cite{breschi2016piecewise,mejari2018towards}, as $\Sigma_{\mathrm{B-PWA}}$;
	
	\item The queuing network described in \cite[Ex.~5]{garbi2020learning}, in which we assume the second and third components of the vector $\mu$ are each restricted to the interval $[10,100]$,
	%
	%
	while its first component remains set to $30$ and the vector $s$ set to $[1000,11,11]$. The system is excited using a random white sequence drawn from a uniform distribution defined on the range above. The resulting ODEs are integrated using the LSODA suite~\cite{hindmarsh1983odepack} via Scipy. The sampling time is set to 5 seconds. We will refer to this system as $\Sigma_{\mathrm{QN}}$.
	
\end{itemize}

The numerical results for the examples above are reported in Table~\ref{tab:PWA_results}, which clearly show how our methodology works well, achieving performance comparable with \gls{PARC} and, in turn, with the current state-of-the-art. 
From our numerical experience we note that solving the training problem in \eqref{eq:training} on, e.g., the $\Sigma_2$ benchmark, required around $15$[s], while PARC approximately $30$[s].

\ifTwoColumn
	\begin{table*}
		\centering
		\begin{tabular}{c|ccc|ccc}
			& {median  BFR} &  {best BFR} & {variance BFR} &{median RMS} & {best RMS} & variance RMS \\
			\hline
			$\Sigma_{\mathrm{PWA}}$& 0.952 (0.937) & 0.959 (0.944) & $2.88 \cdot 10^{-4}$ ($2.66 \cdot 10^{-5}$) & 0.048 (0.064)  &  0.042 (0.057) &   $2.95 \cdot 10^{-4}$ ($3.12 \cdot 10^{-5}$) \\
			$\Sigma_{\mathrm{B-PWA}}$ & 0.883 (0.855)& 0.932 (0.891) &  $7.28 \cdot 10^{-3}$ ($7.48 \cdot 10^{-4}$)& 0.084 (0.109) & 0.050 (0.082) &  $4.20 \cdot 10^{-3}$ ($4.58 \cdot 10^{-4}$) \\
			$\Sigma_{\mathrm{QN}}$& 0.838 (0.81)& 0.851 (0.829) & $2.20 \cdot 10^{-4}$ ($4.41 \cdot 10^{-4}$)& 0.193 (0.224) & 0.175 (0.205)&  $3.40 \cdot 10^{-4}$ ($5.90 \cdot 10^{-4}$) \\
		\end{tabular}
		\caption{BFR and RMS obtained by our method and \gls{PARC} (within parentheses) on \gls{PWA} benchmarks.}
		\label{tab:PWA_results}
	\end{table*}
\else
	\begin{table*}
		\centering
		\begin{tabular}{c|ccc}
			& {median  BFR} &  {best BFR} & {variance BFR} \\
			\hline
			$\Sigma_{\mathrm{PWA}}$& 0.952 (0.937) & 0.959 (0.944) & $2.88 \cdot 10^{-4}$ ($2.66 \cdot 10^{-5}$)\\
			$\Sigma_{\mathrm{B-PWA}}$ & 0.883 (0.855)& 0.932 (0.891) &  $7.28 \cdot 10^{-3}$ ($7.48 \cdot 10^{-4}$)\\
			$\Sigma_{\mathrm{QN}}$& 0.838 (0.81)& 0.851 (0.829) & $2.20 \cdot 10^{-4}$ ($4.41 \cdot 10^{-4}$)\\
		\end{tabular}
		
		\begin{tabular}{c|ccc}
			&{median RMS} & {best RMS} & variance RMS\\
			\hline
			$\Sigma_{\mathrm{PWA}}$& 0.048 (0.064)  &  0.042 (0.057) &   $2.95 \cdot 10^{-4}$ ($3.12 \cdot 10^{-5}$) \\
			$\Sigma_{\mathrm{B-PWA}}$& 0.084 (0.109) & 0.050 (0.082) &  $4.20 \cdot 10^{-3}$ ($4.58 \cdot 10^{-4}$) \\
			$\Sigma_{\mathrm{QN}}$& 0.193 (0.224) & 0.175 (0.205)&  $3.40 \cdot 10^{-4}$ ($5.90 \cdot 10^{-4}$)\\
		\end{tabular}
		\caption{BFR and RMS obtained by our method and \gls{PARC} (within parentheses) on \gls{PWA} benchmarks.}
		\label{tab:PWA_results}
	\end{table*}
\fi

\subsubsection{Linear parameter-varying and nonlinear benchmarks}
\label{sec:NL-bench}
We now compare the performance of our \gls{NN}-based approach and \gls{PARC} on the following nonlinear or parameter-varying system models:
\begin{itemize}   
	
	\item The two-tank system available with the MATLAB system identification toolbox \cite{mathworksControlTwoTank}. 
	As the dataset contains input-output data only, we define the state as $x_t=[y_{t-2} \; y_{t-1} \; y_{t}]^\top$. 
	For this example, we use 2000 samples for training and 1000 for validation without adding any white noise to the measurements.  We will refer to this system as $\Sigma_{\mathrm{Tank}}$;
	
	\item The linear parameter-varying dynamics in \cite[\S IV.B]{breschi2016identification}, excited according to the related discussion in that paper. We will refer to this system, which has also been used as benchmark in~\cite{breschi2016piecewise,mejari2018towards}, as $\Sigma_{\mathrm{B-LPV}}$;
	
	\item The tank system ``$\Sigma_2$'' described in \cite{masti2021learning}, which exhibits a strong nonlinear input-output behaviour. For this example, the noise treatment and system excitation have been performed as in \cite{masti2021learning}. 
	
\end{itemize}

Table~\ref{tab:NL_results} reports the numerical values obtained for these examples, where we can see that, on the $\Sigma_{\mathrm{B-LPV}}$ benchmark, our method greatly outperforms \gls{PARC}.  On the reference hardware, for $\Sigma_\mathrm{tank}$ PARC takes $25$[s], while our training procedure around $40$[s]. On this benchmark, while the identification performance of our approach is comparable to that of \cite{masti2021learning}, our training time is significantly lower than that reported in \cite{masti2021learning} (around $20$ minutes). This is mostly related to the fact that our method needs to learn a few hundreds of coefficients only.
Finally, the results on $\Sigma_2$ are also very competitive \gls{wrt} those in \cite{masti2021learning}, especially in view of the larger dataset employed in the latter paper (i.e., 20000 against 5000 samples). 

\ifTwoColumn
	\begin{table*}
		\centering
	\begin{tabular}{c|ccc|ccc}
		& {median  BFR} &  {best BFR} & {variance BFR} &{median RMS} & {best RMS} & variance RMS \\
			\hline
			$\Sigma_{\mathrm{Tank}}$& 0.901 (0.900) & 0.932 (0.909) & $4 \cdot 10^{-4}$ ($1.64 \cdot 10^{-3}$) & 0.108 (0.109)  &  0.075 (0.09) &  $4.76 \cdot 10^{-4}$ ($1.95 \cdot 10^{-3}$)  \\
			$\Sigma_{\mathrm{B-PLV}}$& 0.696 (0.511)& 0.731 (0.528) &  $4.47 \cdot 10^{-4}$ ($1.62\cdot 10^{-4}$) & 0.287 (0.46) & 0.255 (0.44) &  $3.96 \cdot 10^{-4}$ ($1.30\cdot 10^{-4}$)\\
			$\Sigma_{2}$& 0.924 (0.921)& 0.947 (0.931) &  $1.69 \cdot 10^{-4}$ ($3.47\cdot 10^{-5}$) & 0.061 (0.063) & 0.042 (0.056) &  $1.09 \cdot 10^{-4}$ ($1.96 \cdot 10^{-5}$) \\
		\end{tabular}
		\caption{BFR and RMS obtained by our method and \gls{PARC} (within parentheses) on the nonlinear and LPV benchmarks.}
		\label{tab:NL_results}
	\end{table*}
\else
	\begin{table*}
		\centering
		\begin{tabular}{c|ccc}
			& {median  BFR} &  {best BFR} & {variance BFR}\\
			\hline
			$\Sigma_{\mathrm{Tank}}$& 0.901 (0.900) & 0.932 (0.909) & $4 \cdot 10^{-4}$ ($1.64 \cdot 10^{-3}$)\\
			$\Sigma_{\mathrm{B-PLV}}$& 0.696 (0.511)& 0.731 (0.528) &  $4.47 \cdot 10^{-4}$ ($1.62\cdot 10^{-4}$)\\
			$\Sigma_{2}$& 0.924 (0.921)& 0.947 (0.931) &  $1.69 \cdot 10^{-4}$ ($3.47\cdot 10^{-5}$)\\
		\end{tabular}
		\begin{tabular}{c|ccc}
			& {median RMS} & {best RMS} & variance RMS \\
			\hline
			$\Sigma_{\mathrm{Tank}}$& 0.108 (0.109)  &  0.075 (0.09) &  $4.76 \cdot 10^{-4}$ ($1.95 \cdot 10^{-3}$)  \\
			$\Sigma_{\mathrm{B-PLV}}$& 0.287 (0.46) & 0.255 (0.44) &  $3.96 \cdot 10^{-4}$ ($1.30\cdot 10^{-4}$)\\
			$\Sigma_{2}$& 0.061 (0.063) & 0.042 (0.056) &  $1.09 \cdot 10^{-4}$ ($1.96 \cdot 10^{-5}$) \\
		\end{tabular}
		\caption{BFR and RMS obtained by our method and \gls{PARC} (within parentheses) on the nonlinear and LPV benchmarks.}
		\label{tab:NL_results}
	\end{table*}
\fi

\subsection{Sensitivity analysis -- parameters $n_r^\alpha$, $n_r^\beta$}\label{sub:sensitivity}
We now investigate the sensitivity of our \gls{NN}-based approach \gls{wrt} the main parameters characterizing our technique, i.e., $n_r^\alpha$ and $n_r^\beta$. Note that also $Q_\alpha$, $Q_\beta$ represent further possible hyperparameters to tune, however, we note that according to both the discussion in \cite{hempel2014inverse,hempel2017strong} and our own observation, the performance of our method is not particularly affected by their values. We therefore omit the related sensitivity analysis for these terms.

We report in Table~\ref{tab:sweep_alfa} the median, worst and best \gls{BFR}  achieved on the $\Sigma_2$ and $\Sigma_{\mathrm{B-PWA}}$ for different values of $n_r^\alpha = n_r^\beta$. 
In general we can observe that, while a higher $n_r^\alpha$ is beneficial to improve the best-case scenario performance, it also makes the training problem more difficult.
This is clearly shown by the obtained median \gls{BFR}, especially when dealing with $\Sigma_2$. 
From our numerical experience, the difficulties of training models featuring higher values of $n_r^\alpha$ (and therefore the capability of representing richer continuous \gls{PWA} mappings) can be often overcome by making use of larger dataset.
These results also stress that, in case the date are fixed in advance, choosing appropriate values for $n_r^\alpha$ and $n_r^\beta$ is key. 

\begin{table*}
	\centering
	\begin{tabular}{c|ccc|ccc}
		& \multicolumn{3}{c|}{ $\Sigma_{\mathrm{B-PWA}}$} & \multicolumn{3}{c}{ $\Sigma_{\mathrm{2}}$} \\
		& median & best & variance &median  & best & variance\\
		\hline
		$n_r^\alpha=2$&0.915&0.943&  $1.06\cdot 10^{-3}$ &0.863&0.863& $1.21\cdot 10^{-3}$\\
		$n_r^\alpha=3$&0.897&0.925&  $3.51\cdot 10^{-4}$ &0.889&0.913 & $8.92\cdot 10^{-5}$\\
		$n_r^\alpha=5$&0.904&0.932&  $3.23\cdot 10^{-3}$ &0.915&0.924 & $7.05\cdot 10^{-5}$\\
		$n_r^\alpha=7$&0.883&0.932&  $7.28\cdot 10^{-3}$ &0.924&0.947& $1.69\cdot 10^{-4}$\\
		$n_r^\alpha=11$&0.845&0.936&  $5.99\cdot 10^{-3}$ &0.939&0.956 &$2.85\cdot 10^{-4}$\\
		$n_r^\alpha=17$&0.875&0.910&  $5.87\cdot 10^{-3}$ &0.946&0.952&$2.49\cdot 10^{-4}$\\

	\end{tabular}
	\caption{BFR obtained by our method in learning the behavior of $\Sigma_\mathrm{2}$ and  $\Sigma_{\mathrm{B-PWA}}$ for different values of $n_r^\alpha$.}
	\label{tab:sweep_alfa}
\end{table*}

%

\subsection{Performance for predictive control}\label{sub:pred_con}


The computational advantages of employing a model as in \eqref{eq:alfa_beta} for optimal control purposes have been already analyzed in \cite{hempel2017strong}. Here, we have shown that the strong stationarity conditions offered by Lemma~\ref{lemma:strong_stationarity} can be solved through standard \gls{NLP} solvers such as, e.g., IPOPT \cite{wachter2006implementation}, outperforming classical mixed-integer programming approaches to hybrid system optimal control.

Nevertheless, one may wonder how the adopted \gls{NN} model can be used within an \gls{OCP} in conjunction with a sensitivity-based \gls{NLP} solver. This is relevant for systems as those introduced in~\S \ref{sec:NL-bench}, which do not exhibit behaviors involving logic and dynamic at the same time. 
We then design a simple \gls{OCP} for $\Sigma_2$, mimicking the setup of~\cite{masti2021learning}, with prediction horizon equal $ T = 7$.  We solve, at each step, the following optimization problem:
$$
	\left.
	\begin{aligned}
		&\underset{u_1,\dots,u_7}{\textrm{min}} & & \sum_{i=1}^{7}(x^{+}_{i,2}-r_{i+1})^2 + 0.01~ u_i^2 + 0.001~\delta_i^2\\
		&~~\textrm{ s.t. } & & \hat x^{+}_i=\alpha_\theta(x_i,u_i)-\beta_\theta(x_i,u_i),~i=1,\dots,7,\\
		&&& \delta_i =u_i-u_{i-1},~i=1,\dots,7, \\
		&&& 0.95 \leq u_i \leq 1.2,~i=1,\dots,7,
	\end{aligned}	
	\right.
$$
where with $x_{i,2}$ we mean the second component of $x_i$.

\begin{figure}
	\centering
	\includegraphics[width=\columnwidth]{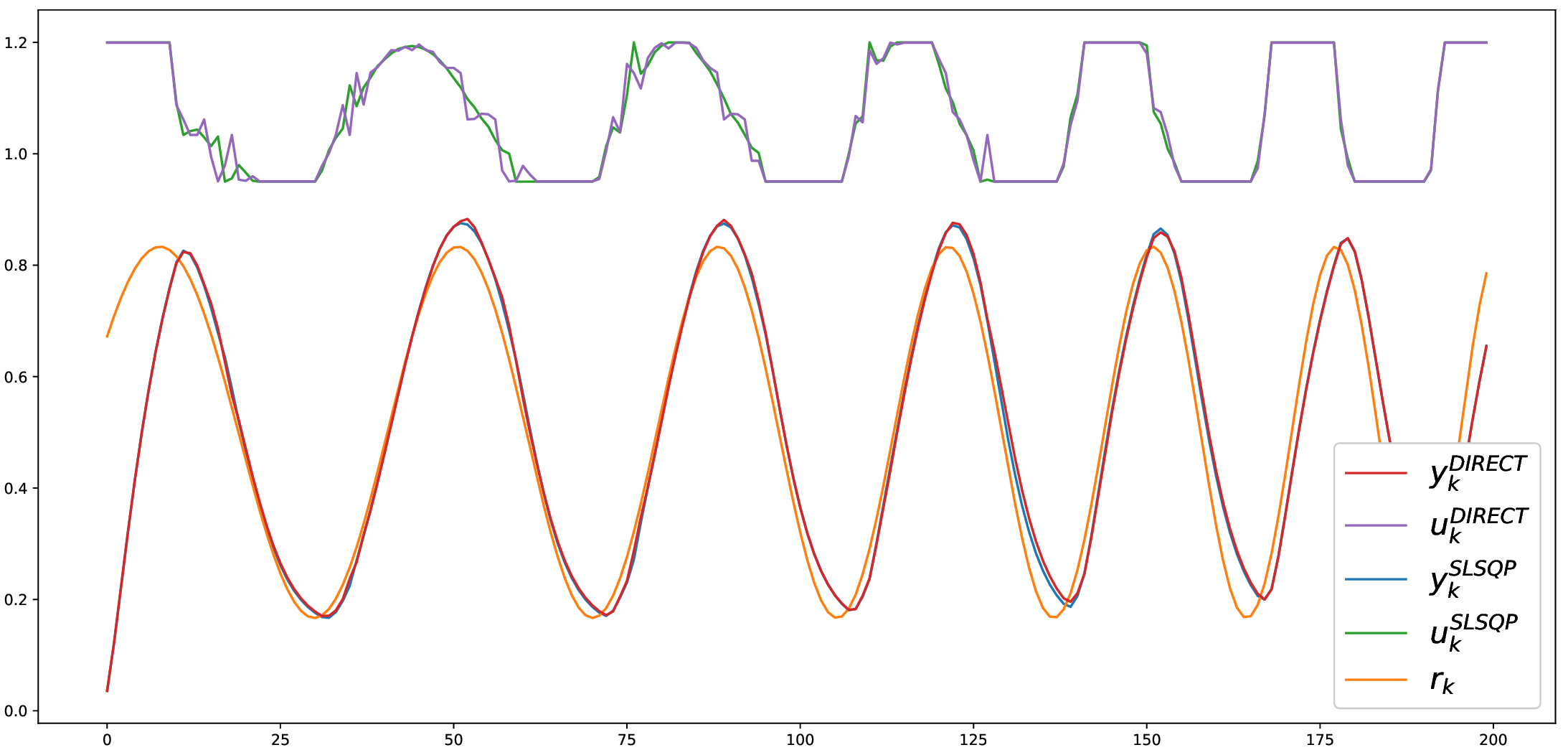}
	\caption{Comparison between an optimal controller based on SQP solver and one based on the DIRECT global optimization solver when tracking the reference sine-sweep $r_k$. The $y$-axis represents the (normalized) magnitude of the quantities involved, while the $x$-axis denotes the time steps. Both controllers have been parametrized to solve the same optimization problem based on a learnt model of $\Sigma_2$.}
	\label{fig:MPC}
\end{figure}

We then compare the behaviour of a closed-loop trajectory achieved when using \texttt{SLSQP} solver with the one obtained with the global optimization solver \texttt{DIRECT\_L}~\cite{gablonsky2001locally}. The results are shown in Fig.~\ref{fig:MPC} where, although the selected input sequence is slightly different, the closed-loop trajectories achieved by the two controllers are almost indistinguishable, thereby showing that derivative-based solver may exploit the adopted \gls{NN}-based models. 
%
%

We note that on our reference hardware, the whole closed-loop simulation with the \texttt{SLSQP}-based controller required approximately one millisecond per time-step.
Overall, the \texttt{SLSQP}-based approach requires around $4$[s], in contrast to the $25$[s] needed by that based on \texttt{DIRECT\_L}.  
While we opted for ``off-the-shelf'' solvers, adopting dedicated methodologies such as, e.g., \cite{Nurkanovic2024MPCC,Fletcher2004jw}, would yield even better computational performance.

\section{Conclusion and outlook}
	We have proposed a \gls{NN}-based methodology to system identification for control that only requires the training of a \gls{NN} via standard tools as simple identification step, and yields a hybrid system model suited for optimal control design. Specifically, we have employed a \gls{NN} with specific, yet simple, architecture, which turns out to be end-to-end trainable and produces a hybrid system with \gls{PWA} dynamics from available data. Extensive numerical simulations have illustrated that, as a \gls{NN}-based identification procedure, our technique has very similar performance compared to the state-of-the-art of hybrid system identification methods. By relying on available results \cite{hempel2017strong}, we have also shown that, under a careful choice of some weights of the \gls{NN}, the resulting hybrid dynamics can be controlled locally optimally by just solving the \gls{KKT} system associated to the underlying finite horizon \gls{OCP}. This is computationally advantageous compared to traditional approaches to optimal control of hybrid systems, usually requiring mixed-integer optimization.
	
	Future work will concentrate on the integration of OptNet layers together with standard explicit layers characterized by different activation functions. In view of the requirement in Corollary~\ref{cor:opt_control_nn}, it is also key to investigate the practical impact the various strategies to impose the conditions in \eqref{eq:strict_inequalities} have on the resulting \gls{NN} model.
	
%
%



\bibliographystyle{plain}
\bibliography{ML_hybrid_systems}



\end{document}